\documentclass[pr,showpacs,preprintnumbers,superscriptaddress,amsmath,amssymb,twocolumn,nobibnotes,longbibliography,nofootinbib]{revtex4-2}%\documentclass[pr,12pt,notitlepage,longbibliography]{revtex4-2}
\usepackage[english]{babel}
\usepackage{mathtools}
\usepackage{graphicx}
\usepackage{mathrsfs}
\usepackage{amssymb}
\usepackage{xcolor}
\usepackage[colorlinks=true,citecolor=blue,urlcolor=blue]{hyperref}
\usepackage{amsmath}
\usepackage{bm}
\usepackage{physics}
\let\div\undefined % Clear the physics package override
\DeclareMathSymbol{\div}{\mathbin}{symbols}{"04} % Restore the standard division sign

\usepackage{nicefrac} 			% diagonal fracs with \nicefrac{}{}

\begin{document}

\title{Electron spin resonance driven photogalvanic effect 	in graphene-based structures
	}
\author{C. Bray}
\altaffiliation{Contributed equally to this work}
\affiliation{L2C UMR 5221, Université de Montpellier, CNRS, 34090 Montpellier, France}

\author{I. Yahniuk}
\altaffiliation{Contributed equally to this work}
\affiliation{Institute of Experimental and Applied Physics, University of Regensburg, 93040 Regensburg, Germany}
\affiliation{CENTERA Labs, Institute of High Pressure Physics, PAS, 01 - 142 Warsaw, Poland}

\author{L. E. Golub}
\affiliation{Institute of Theoretical Physics and Halle-Berlin-Regensburg Cluster of Excellence CCE, University of Regensburg, 93040 Regensburg, Germany}

\author{M. Marocko}
\affiliation{Institute of Experimental and Applied Physics, University of Regensburg, 93040 Regensburg, Germany}

\author{C. Consejo}
\affiliation{L2C UMR 5221, Université de Montpellier, CNRS, 34090 Montpellier, France}

\author{B. Benhamou-Bui}
\affiliation{L2C UMR 5221, Université de Montpellier, CNRS, 34090 Montpellier, France}

\author{Ziyang Gan}
\affiliation{Institute of Physical Chemistry, Friedrich Schiller University Jena, 07743 Jena, Germany}
\author{%Antony 
	A. George}
\affiliation{Institute of Physical Chemistry, Friedrich Schiller University Jena, 07743 Jena, Germany}
\author{%Andrey 
	A.~Turchanin}
\affiliation{Institute of Physical Chemistry, Friedrich Schiller University Jena, 07743 Jena, Germany}

\author{P.~Sadovyi}
\affiliation{Institute of High Pressure Physics of the Polish Academy of Sciences, 01-142 Warsaw, Poland}

\author{K. Watanabe}
\affiliation{Research Center for Electronic and Optical Materials, National Institute for Materials Science, 1-1 Namiki, Tsukuba 305-0044, Japan}

\author{T. Taniguchi}
\affiliation{Research Center for Materials Nanoarchitectonics,  National Institute for Materials Science, 1-1 Namiki, Tsukuba 305-0044, Japan }

\author{J. Eroms}
\affiliation{Institute of Experimental and Applied Physics, University of Regensburg, 93040 Regensburg, Germany}

%\author{J. Wunderlich}
%\affiliation{Physics Department, University of Regensburg, 93040 Regensburg, Germany}
\author{J. Fabian}
\affiliation{Institute of Theoretical Physics and Halle-Berlin-Regensburg Cluster of Excellence CCE, University of Regensburg, 93040 Regensburg, Germany}

\author{F. Teppe}
\affiliation{L2C UMR 5221, Université de Montpellier, CNRS, 34090 Montpellier, France}

\author{S. D. Ganichev}
\affiliation{Institute of Experimental and Applied Physics, University of Regensburg, 93040 Regensburg, Germany}
\affiliation{CENTERA Labs, Institute of High Pressure Physics, PAS, 01 - 142 Warsaw, Poland}
\email{sergey.ganichev@ur.de}

		\begin{abstract}
We report an electron-spin-resonance-driven linear photogalvanic effect (LPGE) in unbiased monolayer graphene and WSe$_2$/graphene heterostructures. Under linearly polarized 45–75 GHz radiation, the photovoltage exhibits pronounced resonant features in both Faraday and Voigt geometries. Multiple resonances associated with the electron spin resonance in graphene are observed for both out-of-plane and in-plane magnetic-field orientations.  Their magnetic-field positions vary linearly with frequency, their amplitudes reverse sign across the charge-neutrality point, and the resonant contribution has the opposite sign to the nonresonant Drude photogalvanic background. We develop a microscopic theory in which radiation-induced momentum alignment followed by skew scattering generates both contributions. Their opposite signs originate from the orthogonal momentum alignments produced by indirect Drude absorption and direct spin-resonant transitions. The theory describes well the main features of the observed resonant photocurrent and provides a microscopic description of ESR-induced LPGE in two-dimensional systems. These results establish the photogalvanic response as a probe of ESR in unbiased micron-scale graphene-based devices.
		\end{abstract}
%	\end{@twocolumnfalse}
	
	\maketitle	
	
	
%	\twocolumn

\section{Introduction}

Electron spin resonance (ESR), also known as electron paramagnetic resonance, is a central spectroscopic tool for probing spin states in condensed matter. The resonance condition $\hbar \omega = g\mu_{\rm B}B$, provides access to the electronic $g$-factor,  while the resonance position, anisotropy, and linewidth contain information on spin-orbit coupling, crystal-field effects, and spin relaxation~\cite{AbragamBleaney1970,PooleFarach1999,WeilBolton2007,SchweigerJeschke2001,Freed2000,Polash2023,Feher1959}. Conventional ESR is generally detected through microwave absorption in a resonant cavity. Although this approach provides high spectral resolution, its sensitivity decreases for micron-scale samples because the signal scales with the number of spins coupled to the cavity mode. Electrical detection circumvents this limitation by monitoring resonance-induced changes in conductivity. The microscopic origin of such signals was later clarified in terms of Rashba spin-orbit splitting~\cite{Rashba1960}, including quantum Hall regime~\cite{Stier2023} and spin-dependent recombination processes, providing a physical basis for electrical detection of spin resonance~\cite{Lepine1972}. Since its early demonstration in semiconductors~\cite{Feher1959,Lepine1972}, electrically detected ESR (EDESR) has therefore become an important and highly sensitive technique for investigating small spin ensembles and device-scale systems~\cite{Stier2023}. In the last decade EDESR has been applied to study various graphene and TMDC/graphene structures~\cite{Mani2012,Lyon2017,Sichau2019,Singh2020,Anlauf2021,Bray2022,Sharma2022,Hild2024a,Morissette2023,Dinar2025}, which have provided access to zero-field splittings associated with energy gaps beyond the Zeeman term. Electrical detection is indeed particularly attractive for two-dimensional materials, whose micron-scale dimensions pose challenges for conventional electromagnetic detection.  However, the applied bias may itself affect the measured response. An unbiased detection scheme based on a dc photocurrent/photovoltage would, therefore, provide a useful alternative for probing spin resonances in micron-scale two-dimensional devices.

\begin{figure}
	\centering
	\includegraphics[width=\linewidth]{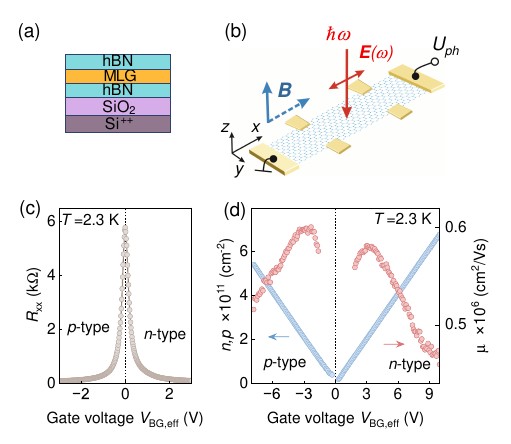}
	\caption{ (a)~Cross-section of the hBN-encapsulated monolayer graphene (hBN/MLG/hBN) device fabricated on a highly doped Si$^{++}$/SiO$_2$ substrate. (b) The Hall-bar structure and the measurement configuration. The downward red arrow represents normally incident, linearly polarized continues wave (cw) GHz radiation with the electric field polarized along the $x$-direction ($\bm E \parallel x$). The orientation of the magnetic field $\bm B$ in the Faraday and Voigt geometries is indicated by the solid and dashed blue arrows, respectively. The photovoltage $U_{\rm ph}$, is measured under unbiased conditions along the Hall-bar structure. (c) Four-terminal longitudinal resistance $R_{xx}$ as a function of the effective back-gate voltage  $V_{\rm BG,eff}$ measured at T = 2.3 K. (d) Carrier density (blue symbols) and mobility (red symbols) against  $V_{\rm BG,eff}$. The vertical dashed lines in panels (c) and (d) indicate the position of the charge neutrality point.
	}
	\label{fig1_Gr_setup}
\end{figure}

\begin{figure*}[t]
	\centering
	\includegraphics[width=\linewidth]{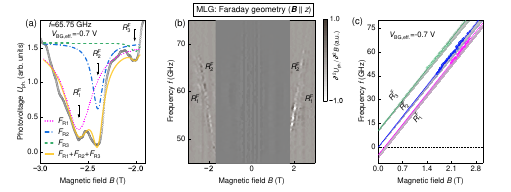}
	\caption{(a):~The photovoltage as a function of the magnetic field is measured in the Faraday geometry $\bm B\parallel z$ in the MLG device. The data were obtained for the radiation with a frequency of 65.75 GHz, an electric field vector parallel to the $x$-axis,  $V_{\rm BG,eff}$ = -0.7~V and a temperature of $T=1.9$~K. The experimental data (gray open circles) are fitted by a sum of three Lorentzian functions (yellow solid curve), with the resonance positions indicated by downward arrows labeled $R_1^{\rm F}$, $R_2^{\rm F}$ and $R_3^{\rm F}$. The corresponding individual Lorentzian are shown by the magenta dotted, blue dashed-dotted and green dashed  curves, respectively. The times used for the fits are calculated using Eq.~\eqref{tau_s} and are 33, 84, and 120.5~ps for the resonances $R_1^{\rm F}$, $R_2^{\rm F}$, and $R_3^{\rm F}$, respectively.  (b):~Color map shows the second derivative of the photovoltage,  $\partial^2 U_{\rm ph} /\partial^2 B$,   with respect to the magnetic field and the frequency of the incoming radiation. The data are presented for $V_{\rm BG,eff}$=-0.7~V and $T=1.9$~K. The semitransparent gray rectangular area in the middle of the map is used to suppress the extremes of the low-B oscillations and highlight the ESR positions. (c):~The magnetic field dependence of the ESR  frequencies is plotted for the resonances $R_1^{\rm F}$ (magenta  circles) and $R_2^{\rm F}$ (blue open circles),  $R_3^{\rm F}$ (green circles). Additional data demonstrating the $R_3^{\rm F}$ resonances  are presented in Fig.~\ref{figA1} of  Appendix~\ref{B}.   Magenta, blue and green solid lines show fits after Eq.~\eqref{fit} with fitting parameters  $g=1.98$ and intercept at $B = 0$ of  $\Delta=-5$, 0 and 10.2~GHz, respectively. The gray areas bordered by dashed lines beneath the data illustrate the range of fitting curves used for the EDESR resonances in graphene structures (Faraday geometry), which were studied in  Refs.~\cite{Mani2012,Lyon2017,Sichau2019,Singh2020,Anlauf2021,Bray2022,Hild2024a,Morissette2023}.
} 
	\label{fig2}
\end{figure*}

\begin{figure*}[t]
	\centering
		\includegraphics[width=\linewidth]{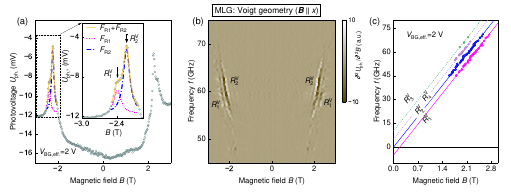}
	\caption{(a):~The photovoltage as a function of the magnetic field is measured in the Voigt configuration (in-plane magnetic field $\bm B \parallel x$) in the MLG device. The data were obtained for the radiation with a frequency of 65~GHz,  an electric field vector parallel to the $x$-axis,  $V_{\rm BG,eff}$ = 2~V and a temperature of $T=1.9$~K. The experimental data (gray circles) are fitted by a sum of two Lorentzian functions (yellow solid curve), with the resonance positions indicated by downward arrows labeled $R_1^{\rm V}$ and $R_2^{\rm V}$. The inset shows the resonance area, which is highlighted by a dashed rectangle. The corresponding individual Lorentzians are shown by the magenta dotted and blue dashed-dotted curves, respectively. The times used for the fits are calculated using Eq.~\eqref{tau_s} and are 85, and 81~ps for the resonances $R_1^{\rm V}$, and $R_2^{\rm V}$, respectively.  (b):~The color map shows the second derivative of the photovoltage,   $\partial^2 U_{\rm ph} /\partial^2 B$, with respect to the magnetic field and the frequency of the incoming radiation. The data are presented for $V_{\rm BG,eff}$= 2~V and $T=1.9$~K.  (c):~The magnetic field dependence of the ESR  frequencies is plotted for the resonances $R_1^{\rm V}$ (magenta  circles), $R_2^{\rm V}$ (blue open circles), $R_3^{\rm V}$ (green open circles) and  $R_4^{\rm V}$ (violet open circles). Additional data demonstrating the resonances $R_3^{\rm V}$ and  $R_4^{\rm V}$ are presented in Fig.~\ref{figA3} of  Appendix~\ref{B}.  Magenta and blue solid  lines show fits of the $R_1^{\rm V}$, and $R_2^{\rm V}$ positions after Eq.~\eqref{fit} with fitting parameters $g=1.98$, and  $\Delta=-5$~GHz and zero, respectively. 		The green and violet dashed lines show the corresponding fits of $R_3^{\rm V}$ and $R_4^{\rm V}$ with fitting parameters of 10.2~GHz and 5~GHz, respectively ($g=1.98$).
	}
	\label{fig3}
\end{figure*}

Here we report the observation of an electron-spin resonance-driven linear photogalvanic effect in unbiased monolayer graphene and WSe$_2$/graphene devices.  Under linearly polarized GHz radiation, both structures exhibit a dc photovoltage with multiple resonances in the Faraday and Voigt geometries, with positions and widths consistent with previous ESR studies after accounting for the renormalization of frequencies and magnetic fields~\cite{Mani2012,Lyon2017,Sichau2019,Singh2020,Anlauf2021,Bray2022,Hild2024a,Morissette2023}. The resonant LPGE has the opposite sign to the nonresonant LPGE background arising from indirect Drude-like optical transitions. The resonance fields vary linearly with radiation frequency, while both the resonant and nonresonant photogalvanic responses reverse sign near the charge-neutrality point. We develop a microscopic theory in which radiation-induced momentum alignment followed by skew scattering generates both the Drude and ESR-driven LPGE currents. The opposite signs of the two contributions result from the orthogonal momentum alignments produced by indirect Drude absorption and direct spin-resonant transitions. It also describes well the generation of the photocurrent in the Voigt configuration. While in general several mechanisms give rise to the Drude-like and ESR LPGE including skew scattering, side-jump and Berry curvature dipole for the Drude like absorption~\cite{Moldavskaya2026} and shift mechanism for the direct transitions, our findings show that in both cases the skew scattering contribution dominates the current.
 
The paper is organized as follows: Sections~\ref{section2} and~\ref{section3} describe the results obtained for MLG and WSe$_2$/MLG 2D materials. These sections are organized similarly. They begin with details of sample preparation and transport characteristics (subsection 1), followed by results on photocurrents detected in Faraday (subsection 2) and Voigt (subsection 3) geometries. Section~\ref{section_Theory} presents the developed theory of the ESR-driven photogalvanic effect and the microscopic model. Section~\ref{Discussion} discusses the experimental results in view of the developed theory. Finally, we summarize the paper and provide an outlook for further research.

\begin{figure*}
	\centering
	\includegraphics[width=\linewidth]{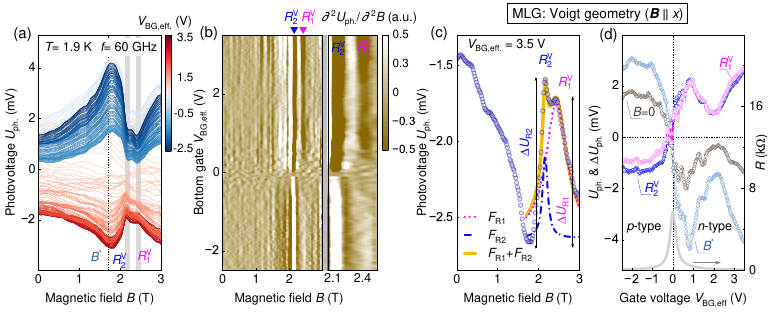}
	\caption{(a):~The photovoltage as a function of the magnetic field is measured in the Voigt configuration (in-plane magnetic field $\bm B \parallel x$) in the MLG device at  effective gate voltages varying  from V$_{\rm BG,eff}=-2.5$ to 3.5~V, see the color bar on the right. The data were obtained for the radiation with a frequency of 60~GHz,  an electric field vector parallel to the $x$-axis, and a temperature of $T=1.9$~K. Vertical dashed lines indicate the positions of the resonances $R_1^{\rm V}$ and $R_2^{\rm V}$, as well as of  the field $B^*=1.7$~T outside of the resonances. (b):~Color map shows the second derivative of the photovoltage,   $\partial^2 U_{\rm ph} /\partial^2 B$, with respect to the magnetic field and the effective bottom gate voltage. The data are presented for the radiation frequency $f=60$~GHz and $T=1.9$~K. 
		%The data for $B$ ranging from zero to 3~T are shown on the left.
	The ESR positions  are highlighted by the downward  triangles and labeled $R_1^{\rm V}$ and $R_2^{\rm V}$, respectively. A zoom of the data highlighting the range of the ESR extrema is shown on the right side separated by the gray bar.  (c): The photovoltage as a function of the magnetic field. The data were obtained for the radiation with a frequency of 60~GHz,  $V_{\rm BG,eff}$ = 3.5~V and a temperature of $T=1.9$~K. The experimental data (blue  circles) are fitted by a sum of two Lorentzian functions (yellow solid curve), with the resonance positions indicated by labels $R_1^{\rm V}$ and $R_2^{\rm V}$. The corresponding individual Lorentzian are shown by the magenta dotted and blue dashed-dotted curves, respectively. The times used for the fits are calculated using Eq.~\eqref{tau_s} and are 18, and 74~ps for the resonances $R_1^{\rm V}$, and  $R_2^{\rm V}$,  respectively.   (d):~The gate voltage dependence  of the photocurrent magnitudes obtained for the magnetic fields $B=0$, $B^*=1.7$~T and those corresponding to the resonances $R_1^{\rm V}=2.42$~T and $R_2^{\rm V}=2.18$~T. The ESR signals $\Delta U_{\rm R1}$ and $\Delta U_{\rm R2}$ were obtained subtracting the Drude contributions at $B^*$, as shown by the double arrows in the panel (c). The  brown open circles show the photocurrent for zero magnetic field.  Gray solid line shows the gate voltage dependence of the two-point resistance and the vertical dashed line indicates the CNP position. 
}
	\label{fig4}
\end{figure*}

	\begin{figure}[!h]
	\centering
	\includegraphics[width=\linewidth]{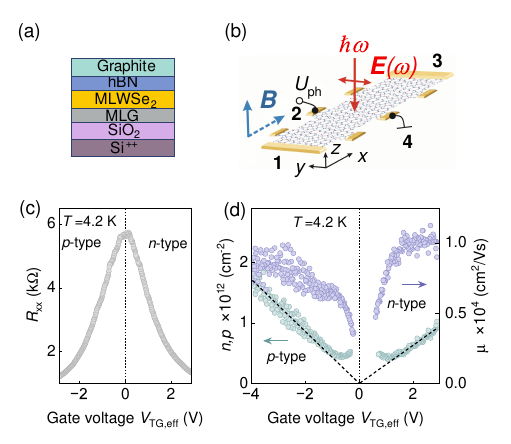}
	\caption{(a):~Cross-section of the WSe$_2$/MLG structure deposited on the SiO$_2$/Si substrate with graphite top gate layer. (b):~Sketch of the Hall bar shaped sample  and the measurement scheme used in the experiments. The downward red arrow illustrates the normally incident, linearly polarized   $cw$ GHz~radiation of the Schottky diode frequency multiplier ($\bm E \perp x$). The solid and dashed blue arrows indicate the direction of the external magnetic field, $\bm B$, used in experiments conducted in Faraday and Voigt configurations, respectively. The photovoltage $U_{\rm ph}$ is measured in an unbiased device perpendicular to the Hall bar structure, i.e., parallel to $y-$direction. (c):~The longitudinal resistance $R_{xx}$ as a function of the effective top gate-gate voltage $V_{\rm TG,eff}$ measured at $V_{\rm BG} = 0$ and $T = 4.2$~K. (d):~Dependence of carrier density  (green symbols) and mobility (violet symbols) on ($V_{\rm TG,eff}$).  The vertical dashed lines in panels (c) and (d) indicate the position of the charge neutrality point. 
	}
	\label{fig5}
\end{figure}

\begin{figure*}
	\centering
	\includegraphics[width=0.85\linewidth]{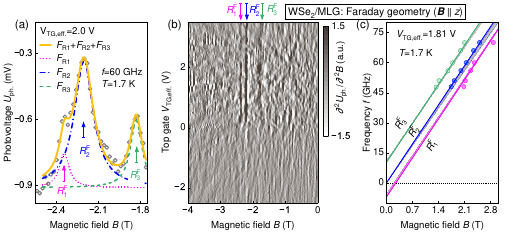}
	\caption {(a):~The photovoltage as a function of the magnetic field is measured in the Faraday geometry $\bm B\parallel z$ in the WSe$_2$/MLG device at  $V_{\rm BG}=30$~V. The data were obtained for the radiation with a frequency of 60~GHz, an electric field vector parallel to the $y$-axis,  $V_{\rm TG,eff} = 1.81$~V and a temperature of $T=1.7$~K. The experimental data  (blue  circles) are fitted by a sum of two Lorentzian functions  (yellow solid curve), with the resonance positions indicated by upward arrows labeled $R_1^{\rm F}$ and $R_2^{\rm F}$. The corresponding individual Lorentzian are shown by the  magenta dotted and blue dashed-dotted curves, respectively. The times used for the fits are calculated using Eq.~\eqref{tau_s} and are 126 and 100~ps for the resonances $R_1^{\rm F}$, and $R_2^{\rm F}$, respectively.  (b):~The color map shows the second derivative of the photovoltage,  $\partial^2 U_{\rm ph} /\partial^2 B$,   with respect to the magnetic field and top gate voltage ranging from  $V_{\rm TG,eff}=-2.5$~V to 3.5~V.  The data are presented for $f=60$~GHz and $T=1.7$~K. The ESR positions are indicated by downward arrows labeled as $R_1^{\rm F}$, $R_2^{\rm F}$ and  $R_3^{\rm F}$. (c):~The magnetic field dependence of the ESR  frequencies is plotted for the resonances $R_1^{\rm F}$  (magenta  circles),   $R_2^{\rm F}$ (blue circles) and $R_3^{\rm F}$ (green circles). The data are obtained for  $V_{\rm TG,eff} = 1.81$~V and a temperature of $T=1.7$~K. Magenta, blue and green solid lines show fits after Eq.~\eqref{fit} with fitting parameters  $g=1.99$ and intercept at $B = 0$ of  $\Delta=-7$, 0 and 10.2~GHz, respectively. The gray areas bordered by dashed lines beneath the data illustrate the range of fitting curves used for the EDESR resonances in graphene structures (Faraday geometry), which were studied in  Refs.~\cite{Sharma2022,Dinar2025}.
}
	\label{fig6}
\end{figure*} 

\begin{figure*}
	\centering
	\includegraphics[width=\linewidth]{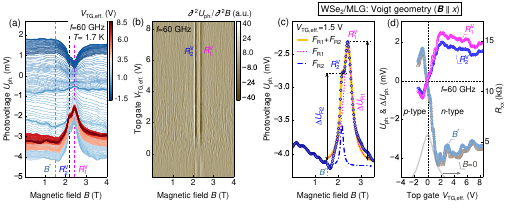}	
	\caption{(a):~The photovoltage as a function of the magnetic field is measured in the Voigt configuration (in-plane magnetic field $\bm B \parallel x$) in the WSe$_2$/MLG device at  effective gate voltages varying  from V$_{\rm TG,eff}=-1.5$ to 8.5~V ($V_{\rm BG}=30$~V), see the color bar on the right. The data were obtained for the radiation with a frequency of 60~GHz,  an electric field vector parallel to the $y$-axis,  a temperature of $T=1.7$~K and back gate voltage 30~V. Vertical dashed lines indicate the positions of the resonances $R_1^{\rm V}$ and $R_2^{\rm V}$, as well as of  the field $B^*=1.5$~T outside of the resonances. (b):~The color map shows the second derivative of the photovoltage,   $\partial^2 U_{\rm ph} /\partial^2 B$, with respect to the magnetic field and the effective top gate voltage. The data are presented for the radiation frequency $f=60$~GHz and $T=1.7$~K. The ESR positions  are  labeled as $R_1^{\rm V}$ and $R_2^{\rm V}$.  (c): The photovoltage as a function of the magnetic field. The data were obtained for the radiation with a frequency of 60~GHz,  $V_{\rm TG,eff} = 1.5$~V and a temperature of $T=1.7$~K. The experimental data  (blue  circles) are fitted by a sum of two Lorentzian functions  (yellow solid curve), with the resonance positions indicated by labels $R_1^{\rm V}$ and $R_2^{\rm V}$. The corresponding individual Lorentzian are shown by the  magenta dotted and blue dashed-dotted curves, respectively. The times used for the fits are calculated using Eq.~\eqref{tau_s} and are 26 and 74~ps for the resonances $R_1^{\rm V}$, and $R_2^{\rm V}$, respectively.   (d):~The gate voltage dependence  of the photocurrent magnitudes obtained for the magnetic fields $B=0$, $B^*=1.5$~T and that corresponding to the resonances $R_1^{\rm V}=2.42$~T and $R_2^{\rm V}=2.18$~T. The ESR signals $\Delta U_{\rm R1}$ and $\Delta U_{\rm R2}$ were obtained subtracting the Drude contributions at $B^*$, as shown by the double arrows in the panel (c). The  brown  circles show the photocurrent for zero magnetic field.  Gray solid line shows the gate voltage dependence of the two-point resistance and the vertical dashed line indicates the CNP position.   
	}
	\label{fig7}
\end{figure*}

\begin{figure*}
	\centering
	\includegraphics[width=\linewidth]{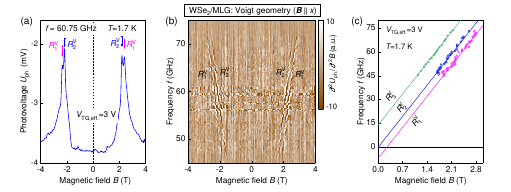}
	\caption{(a):~The photovoltage as a function of the magnetic field is measured in the Voigt configuration (in-plane magnetic field $\bm B \parallel x$) in the WSe$_2$/MLG device. The data were obtained for the radiation with a frequency of 60.75~GHz,  an electric field vector parallel to the $y$-axis,  $V_{\rm BG,eff}$ = 3~V ($V_{\rm BG} = 30$~V), a temperature of $T=1.7$~K and back gate voltage of 30~V. The downward arrows labeled as indicate the resonance positions labeled as $R_1^{\rm V}$ and $R_2^{\rm V}$. The times used for the Lorentzian fits are calculated using Eq.~\eqref{tau_s} and are 26 and 97~ps for the resonances $R_1^{\rm V}$, and $R_2^{\rm V}$, respectively.  (b):~The color map shows the second derivative of the photovoltage,   $\partial^2 U_{\rm ph} /\partial^2 B$, with respect to the magnetic field and the frequency of the incoming radiation. The data are presented for $V_{\rm TG,eff}$= 3~V and $T=1.7$~K.  (c):~The magnetic field dependence of the ESR  frequencies is plotted for the resonances $R_1^{\rm V}$  (magenta circles), $R_2^{\rm V}$ (blue circles) and $R_3^{\rm V}$  (green circles). Additional data demonstrating the resonance $R_3^{\rm V}$  are presented in Fig.~\ref{figA2} of Appendix~\ref{B}.	Magenta, blue and green solid lines show fits after Eq.~\eqref{fit} with fitting parameters  $g=1.99$ and intercept at $B = 0$ of  $\Delta=-7$, 0 and 14~GHz, respectively.  	
}
	\label{fig8}
\end{figure*} 

\section{Results for the monolayer graphene structures}
\label{section2}

\textbf{MLG samples}:
The exfoliated monolayer graphene is encapsulated between bottom and top hBN layers, each with a thickness of approximately 30~nm. Then, the hBN/MLG/hBN structure was deposited onto a   Si$^{++}$/SiO$_2$ substrate. Highly doped silicon Si$^{++}$ serves as an effective back gate and a 285-nm-thick layer of silicon dioxide (SiO$_2$)  is used as an insulator layer.  Schematic cross-section view of the hBN/MLG/hBN structure on Si$^{++}$/SiO$_2$ is shown in Fig.~\ref{fig1_Gr_setup}(a). The device was patterned into a Hall-bar geometry with a channel length of approximately 20~$\mu$m (defined as the distance between the source and drain contacts), as schematically illustrated in Fig.~\ref{fig1_Gr_setup}(b). The Hall-bar structure was fabricated using electron-beam lithography, reactive ion etching, and standard metal deposition techniques. Details of the fabrication process are provided in Ref.~\cite{Sadovyi2025a}.

Sweeping the back gate voltage $V_{\rm BG}$  shows that the charge neutrality point (CNP) is well observable in the longitudinal resistance. For different sample cooldowns the CNP position $V_{\rm CNP}\approx 2.5$~V slightly shifts. Thus, to compare various measurements we use the effective gate voltage $V_{\rm BG,eff} = V_{\rm BG} - V_{\rm CNP}$.  Figure.~\ref{fig1_Gr_setup}(c) and (d) show the dependencies of the four-terminal resistance, carrier density, and carrier mobility on effective back-gate voltage. The device exhibits a high carrier mobility of up to $\mu \approx 0.6 \times 10^6$~cm$^2$/V~s,  confirming the high quality of the investigated structure. The type of carriers and  the dependence of the carrier density on gate voltages ($n=0.68\times 10^{11}V_{\rm BG,eff}$~cm$^{-2}$V$^{-1}$) are defined from Hall measurements. More information on the magnetotransport studies can be found in the supplemental materials to Ref.~\cite{Yahniuk2026}.

\textbf{Methods:} The MLG  Hall-bar samples were mounted in a temperature-regulated, 6~T, horizontal, cryogen-free magnet system with optical access. The system is equipped with two polytetrafluoroethylene (PTFE) windows and a diamond window (cold window). A sub-THz source generated by a Schottky diode with a multiplied frequency was used to obtain linearly polarized radiation with frequencies ranging from 45 to 75~GHz (with an optical power of up to 150 mW). The sub-THz beam is focused on the sample by a gold coated parabolic mirror. The diameter of the THz spot (3-6 mm) was much larger than the sample size, ensuring uniform irradiation.  Radiation was applied at normal incidence with the electric field $\bm E$ oriented parallel to the source drain line of the Hall bar, see Fig.~\ref{fig1_Gr_setup}(b). No bias was applied to the samples for all measurements described below. The signal was detected as a voltage drop from a source-drain pair of ohmic contacts to the Hall-bar samples. The voltage was then amplified and measured via a standard lock-in technique using a Stanford Research SR 860.  The magnetic field $\bm B$ was oriented either perpendicular to the sample's surface (Faraday geometry) or applied in-plane along the Hall bar (Voigt geometry), see blue solid and dashed arrows in Fig.~\ref{fig1_Gr_setup}(b), respectively. 

\textbf{Results obtained in the Faraday configuration}: Figure~\ref{fig2} shows the magnetic field dependence of the resonant photocurrent measured in the Faraday geometry. Figure~\ref{fig2}(a) shows the exemplary behavior of the photocurrent measured in the MLG device at $T=1.9$~K, $V_{\rm BG,eff} = -0.7$~V and a radiation frequency $f=65.75$~GHz. The figure demonstrates that the photocurrent exhibits two resonance dips $R_1^{\rm F}$ and $R_2^{\rm F}$ with the minima at magnetic fields  about -2.6 and -2.41~T, respectively.  An additional resonance $R_3^{\rm F}$ with a substantially smaller magnitude is detected at a magnetic field of about 2~T, see also Fig.~\ref{figA1} in Appendix~\ref{B}.   The resonant dips can be accurately described by Lorentzian functions, see  Fig.~\ref{fig2}(a). From the Half Width at Half Maximum (HWHM), $\Delta B_{\rm HWHM}$, we obtain the spin relaxation time $\tau_s$ after~\cite{AbragamBleaney1970}
\begin{equation}
\label{tau_s}
\tau_s = {\hbar \over g\mu_{\rm B}\Delta B_{\rm HWHM}}.
\end{equation}
The times are 33, 84, and 120.5~ps for the resonances $R_1^{\rm F}$, $R_2^{\rm F}$, and $R_3^{\rm F}$, respectively. The photocurrent dips are observed for both polarities of the magnetic field. Varying the frequency from 45 to 70~GHz we detected that the magnetic field positions of all resonances vary linearly with $f$, see Fig.~\ref{fig2}(b) and (c). The slope of these lines is the same, whereas the intercepts at $B = 0$ are different being   $\Delta=-5$, 0 and 10.2~GHz  for the resonances $R_1^{\rm F}$, $R_2^{\rm F}$, and $R_3^{\rm F}$, respectively.

\textbf{Results obtained in the Voigt configuration}: Figure~\ref{fig3} shows the magnetic field dependence of the photocurrent detected in the Voigt geometry. Alike in the Faraday geometry we observed two resonance dips labeled $R_1^{\rm V}$ and $R_2^{\rm V}$, see Fig.~\ref{fig3}(a). We also detected weak resonant signals, $R_3^{\rm V}$ and $R_4^{\rm V}$, under certain conditions, see  Fig.~\ref{figA3} of the Appendix~\ref{B}. We emphasize that, in both geometries, the resonant photocurrents have an opposite sign from the non-resonant photocurrents and the current has the same sign for positive and negative magnetic field polarity. The resonant dips can be well fitted by Lorentzian functions, see Figs.~\ref{fig3}(a) and~\ref{fig4}(c). Using Eq.~\eqref{tau_s} we extract the spin relaxation times $\tau_s=85$ and 81~ps for $R_1^{\rm V}$ and $R_2^{\rm V}$ resonances, respectively. Figures~\ref{fig3}(b) and (c) show that the position of the resonances scale linearly with the radiation frequency. The lines have the same slope and the intercept at $B = 0$ with values  $\Delta=-5$~GHz ( $R_1^{\rm V}$), zero  ( $R_2^{\rm V}$) 10.2~GHz ($R_3^{\rm V}$) and 5~GHz ($R_4^{\rm V}$). Figure~\ref{fig4} presents the gate voltage dependence of the photocurrent. It reveals that the  resonance position is almost independent of the gate voltage, and that the sign of both the non-resonant and resonant photocurrents is inverted in the vicinity of the CNP, see  Fig.~\ref{fig4}(b) and (d). To extract the magnitude of the resonance photocurrents, we subtracted the current magnitude at magnetic field $B^*$, at which the signal starts to decrease, from the total current at the resonance position, see Fig.~\ref{fig4}(c).

\section{Results for the WSe$_2$/MLG 2D material}
\label{section3}

\textbf{WSe$_2$/MLG 2D material:}
%Marina Marocco: Sample G-41!
 The WSe$_2$/MLG 2D materials were prepared as follows. Graphene and hBN flakes were exfoliated from bulk material on Si chips covered with 90 nm thermal SiO$_2$. Suitable flakes were identified and a van der Waals heterostructure was assembled following the procedure detailed by Purdie et al.~\cite{Purdie2018}. Figure~\ref{fig5}(a) shows a cross section of the fabricated device.  First, a 20~nm thick hBN flake was picked up using a polycarbonate film on a dome-shaped polydimethylsiloxane droplet. CVD-grown WSe$_2$~\cite{George2019,Turchanin2024} and single layer graphene were subsequently picked up, aligning fractured crystal edges with approximately $15^\circ$  twist angle for maximum spin-orbit coupling~\cite{Rockinger2026} and avoiding the ambiguity of zigzag and armchair edges. The final stack was transferred to Si covered with 285~nm chlorinated thermal oxide. The described stacks were prepared as Hall bar structures, see Fig.~\ref{fig5}(a). Contacts were defined by electron beam lithography and evaporation of 0.5 nm Cr and 30 nm Au (fine wiring) or 0.5 nm Cr, and 100~nm Au (bond pads), and the mesa was etched by reactive ion etching using O$_2$ for graphene and SF$_6$ for hBN and WSe$_2$~\cite{Pizzocchero2016}. Top and bottom gates were used to vary the carrier density and structure asymmetry. For the top gate, we placed a 21~nm thick hBN flake on top of the finished stack to provide gate insulation, and then transferred a few layer graphene flake, connecting to predefined Cr/Au contact leads. The highly doped Si substrate is used as a back gate, contacted by silver paste to a chip carrier, and pads are wire bonded with aluminum.  Figure~\ref{fig5}(c) shows the transport results obtained by applying a current of 50~nA and measuring the voltage on the edge longitudinal contacts.  The four-probe resistance, $R_{xx}$, as a function of the effective top gate voltage, $V_{\rm TG,eff} = V_{\rm TG} - V_{\rm CNP}$, is measured at $V_{\rm BG} = 0$ and $T=4.2$~K; see the gray circles in Fig.~\ref{fig5}(c).  A pronounced peak at $V_{\rm TG,eff} = 0$ corresponds to the CNP, where $V_{\rm TG}$ is the applied top gate voltage and $V_{\rm CNP} \approx 2$~V. Note that while the results presented below were obtained at different values of the back gate voltage and, consequently, $V_{\rm CNP}$, the  $V_{\rm TG,eff} =0$ always corresponds to the CNP.

Figure~\ref{fig5}(d) shows the dependencies of the  carrier density and carrier mobility on effective back-gate voltage. The device exhibits a  carrier mobility of up $\mu \leq 1 \times 10^4$~cm$^2$/V~s. The type of carriers and  the dependence of the carrier density on gate voltages at zero back gate voltage ($n=0.68\times 10^{11}V_{\rm BG,eff}$~cm$^{-2}$V$^{-1}$ ) are defined from Hall measurements. Figure.~\ref{fig1_Gr_setup}(c) and (d) show the dependencies of the four-terminal resistance, carrier density, and carrier mobility on effective back-gate voltage. More information on the magnetotransport studies of similar structure fabricated in our lab following the same recipe can be found in Ref.~\cite{Rockinger2026}.

\textbf{Methods:} The unbiased WSe$_2$ Hall-bar samples were studied in the same set-up as described in the paragraph ''methods'' of Sec.~\ref{section2}. Figure~\ref{fig5}(b) shows the orientations of the magnetic field and the radiation electric field $\bm E$ used in the measurements presented below. The signal was detected as a voltage drop from a  pair of ohmic contacts   on opposite sides of the Hall-bar (contacts 2 and 4) and the radiation electric field $\bm E$ was also oriented perpendicular to the line connecting the source-drain contacts, see Fig.~\ref{fig5}(b).  The magnetic field $\bm B$ was oriented either perpendicular to the sample's surface (Faraday geometry) or applied in-plane along the Hall bar (Voigt geometry), see blue solid and dashed arrows in Fig.~\ref{fig5}(b), respectively. 

\textbf{Results obtained in the Faraday configuration}: 
Figure~\ref{fig6} shows the magnetic field dependence of the photocurrent excited in WSe$_2$/MLG at $T=1.7$~K irradiated by $f=60$~GHz. Also in this structure we detected two strong resonant dips, which can be well described by the Lorentzian functions, see Figure~\ref{fig6}(a). Using Eq.~\eqref{tau_s} we extract the spin relaxation times $\tau_s=126$ and 100~ps for $R_1^{\rm F}$ and $R_2^{\rm F}$ resonances, respectively. The resonance position does not depend on the gate voltage, and the sign of the resonant signals reverses at the CNP, see Fig.~\ref{fig6}(c). Additionally, we detected a weak third resonance in some experiments, as seen in panels (a) and (c). The resonances scale linearly with the radiation frequency, see Fig.~\ref{fig6}(c), the slope of the lines is the same. The lines for the resonances $R_1^{\rm F}$, $R_2^{\rm F}$, and $R_3^{\rm F}$  have intercepts at $B = 0$ at   $\Delta=-7$, 0 and 10.2~GHz, respectively.

\textbf{Results obtained in the Voigt configuration}: The resonance dips are also detected in the Voigt configuration. Figure~\ref{fig7} shows magnetic field and top gate dependencies of the photocurrent excited by the radiation with frequency 60~GHz. As in the experiments described above the resonant photocurrent has consistently opposite sign to the nonresonant and zero-magnetic field photocurrents. The resonances  labeled $R_1^{\rm V}$ and $R_2^{\rm V}$  are well fitted by the Lorentzian function, see Fig.~\ref{fig7}(c). The spin relaxation times are $\tau_s=26$ and 74~ps for $R_1^{\rm V}$ and $R_2^{\rm V}$ resonances, respectively.  While the resonance positions are independent of the gate voltage, see Fig.~\ref{fig6}(a) and (b), all individual resonant and non-resonant photocurrent contributions change their sign in the vicinity of the CNP, see Fig.~\ref{fig7}(d). Figure~\ref{fig8}(a) shows that the signs of both the resonant and nonresonant photocurrents are the same for positive and negative magnetic fields. A weak third resonance $R_3^{\rm V}$ is also present in the Voigt geometry, see  Fig.~\ref{figA2} of the Appendix~\ref{B}.  The positions of the resonances scale linearly with radiation frequency, see Figs.~\ref{fig8}(b) and~(c). The slope of the line is the same for all resonances and the intercepts at $B=0$ are  $\Delta = -7$~GHz ($R_1^{\rm V}$), 0 ($R_2^{\rm V}$) and 14~GHz ($R_3^{\rm V}$).

\section{Theory}
\label{section_Theory}

\subsection{Photogalvanics in graphene at Drude absorption}

\begin{figure}[!h]
	\centering
	\includegraphics[width=\linewidth]{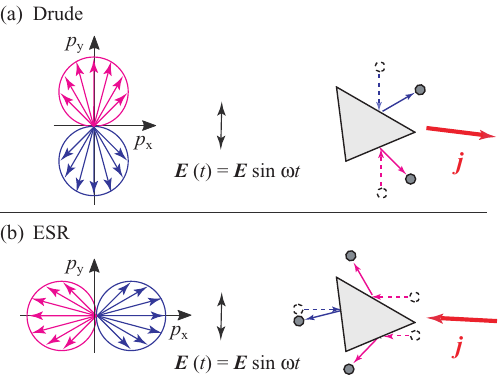}
	\caption{Microscopic mechanism of the  photocurrent formation at Drude (panel a) and ESR (panel b) absorption. Upon absorption of linearly-polarized radiation, an alignment of electron momenta takes place: there is a larger number of particles with momenta along and opposite to the electric field direction and a smaller number in the perpendicular direction (at energy $\varepsilon_p > \varepsilon_{\rm F}$). The corresponding stationary momentum distribution is described by Eqs.~\eqref{f2},~\eqref{align_ESR} and is sketched on the left sides of the panels (a) and (b). Scattering of these momentum-aligned carriers (dashed arrows) by asymmetrical scatterers (shown by irregular triangles) results in uncompensated charge flows (solid arrows) and in a dc electric current (red arrows). For Drude (indirect optical transitions) and ESR (direct optical transitions) absorption, the momentum alignment takes place for the directions along and perpendicular to the radiation electric field, respectively. This results in the opposite photocurrent directions, as illustrated on the right sides of the panels (a) and (b).  
	}
	\label{fig9}
\end{figure}

Since the experiments show the photocurrents at normal incidence, we conclude that our structures have the lowest possible symmetry without any reflection planes or rotation axes. The corresponding point symmetry group is $C_1$. In the GHz frequency range for a sufficiently doped monolayer graphene, the radiation absorption is caused by intraband, Drude-like, absorption. The photocurrent at low temperatures is formed due to elastic skew scattering by asymmetrical defects which is allowed by the $C_1$ symmetry. The necessary ingredient for the skew-scattering induced photocurrent is {\em momentum alignment} occurring at the radiation absorption. This is a formation of the anisotropic dc distribution in momentum space in the second order in the radiation electric field amplitude and described by the second angular harmonics, see Fig.~\ref{fig9}. This distribution is even in momentum, so it does not lead to the electric current. However, skew scattering of momentum-aligned carriers results in the photogalvanic current. Note that in the discussed experiments the edge photocurrent can be excluded, because it vanishes for the electric field oriented along or perpendicular to the edges~\cite{Candussio2021,Candussio2021a}. Figure~\ref{fig9} illustrates the photocurrent formation. It is shown in Fig.~\ref{fig9}(a) that carriers with $p_y>0$ (see magenta solid and dashed arrows) are scattered off to the right semiplane (magenta solid arrow). The carriers with $p_y<0$ (see blue solid and dashed arrows) are scattered off also to the right semiplane (blue solid arrow). Thus, the photocurrent is formed, see thick solid red arrow.

The alignment of electron momenta at Drude absorption for carrier energies above the Fermi level is along the electric field. This means that the number of carriers with momenta along and opposite to the radiation electric field $\bm E$ is higher than in the perpendicular directions, see left side of Fig.~\ref{fig9}(a). In the kinetic theory, the photocurrent is obtained by iterating the Boltzmann kinetic equation up to the second order in the electric field $\bm E(t)=\bm E \exp (-i\omega t) + c.c.$~\cite{Moench2024}. The stationary correction to the charge carrier's distribution function describing an alignment of their momenta in graphene  has the following form
\begin{equation}
	\label{f2}
	f_{\bm p}^\text{(al)} =  |\bm E|^2   {e^2 v^2\tau_2 \varepsilon_p\over 2} \qty({\tau_1 f_0'/\varepsilon_p\over 1+ \omega^2\tau_1^2}  )' \cos{(2\varphi_{\bm p}-2\alpha)},
\end{equation}
where $\alpha$ is an angle between the direction of the $\bm E$ and $x$ axis,  $\bm p$ and $\varphi_{\bm p}$ are 2D momentum and its polar angle, $v$ is the Dirac fermion velocity, $\varepsilon_p=vp$ is the carrier energy, $f_0$ is the Fermi-Dirac distribution function, prime means differentiation over $\varepsilon_p$, and $\tau_1$ and $\tau_2$ are the energy-dependent relaxation times  of the first and second harmonics of the distribution function. They are given by $\tau_n^{-1}(\varepsilon_p)=\sum_{\bm p'}W^s_{\bm p' \bm p} [1-\cos{n(\varphi_{\bm p'}-\varphi_{\bm p})}]$, where $W^s_{\bm p' \bm p}=W^s_{\bm p \bm p'}$ is the symmetrical part of the elastic scattering probability.

Skew scattering of the momentum-aligned carriers results in the photogalvanic current. This is shown on the right side of  Fig.~\ref{fig9}(a). Generalization of the results for the photocurrent $\bm j^{\rm Drude}$ driven by linearly polarized radiation in structures with $C_1$ symmetry~\cite{Otteneder2020} and in systems with linear energy dispersion~\cite{Olbrich2014} yields for the components in the graphene plane
\begin{align}
	\label{j_Drude}
	&j_{x,y}^{\rm Drude} = 
	\mathcal P_{\rm Drude}
	\\ &\times 
	\biggl\{
	{\left[\varepsilon_\text{F}^2 \tau_{\rm al} (\Xi_{c,s}P_{\rm L1}+\Lambda_{c,s}P_{\rm L2})\right]'\over \varepsilon_\text{F}^2} 
	+ {\tau_{\rm al}\varepsilon_\text{F}\over \tau_{\rm tr}} \qty({\tau_{\rm tr} \over \varepsilon_\text{F}})' 
	\nonumber
	\\ 
	& \times {1-\omega^2\tau_{\rm tr}\tau_{\rm al}\over 1+\omega^2\tau_{\rm al}^2}[(\Xi_{c,s}\mp\Lambda_{s,c})P_{\rm L1} \mp (\Xi_{s,c}\pm\Lambda_{c,s})P_{\rm L2}]
	\biggr\}. 
	\nonumber
\end{align} 
Here 
\begin{equation}
\mathcal P_{\rm Drude}={2|\bm E|^2 n (ev)^2 \tau_{\rm tr}/\varepsilon_\text{F} \over  1+\omega^2\tau_{\rm tr}^2}
\end{equation}
is the absorbed power density, $P_{\rm L1}=(E_x^2-E_y^2)/|\bm E|^2$ and $P_{\rm L2}=2E_xE_y/|\bm E|^2$ are the Stokes parameters describing the linear polarization state of the radiation, prime here means differentiation over the Fermi energy $\varepsilon_\text{F}$, and $\tau_{\rm tr}$ and $\tau_{\rm al}$ are, respectively, the times $\tau_1$ and $\tau_2$ at ${\varepsilon_p=\varepsilon_\text{F}}$. Note that the derivatives appear due to an opposite sign of contributions to the photocurrent from the states above and below the Fermi energy.

The asymmetry of the scattering probability present in the $C_1$ point group is taken into account by the dimensionless skew factors $\Xi_{c,s}, \Lambda_{c,s} \ll 1$:
\begin{align}
	\label{Xi_Lambda}
	&\Xi_{c,s} = \tau_\text{tr} \sum_{\bm p'}\left< \cos(\sin){\varphi_{\bm p}} W^{a}_{\bm p' \bm p} \cos{2\varphi_{\bm p'}}\right>_{\varphi_{\bm p}},
	\\
	&\Lambda_{c,s} = \tau_\text{tr} \sum_{\bm p'}\left< \cos(\sin){\varphi_{\bm p}} W^{a}_{\bm p' \bm p} \sin{2\varphi_{\bm p'}}\right>_{\varphi_{\bm p}}, \nonumber
\end{align}
where $W_{\bm p \bm p'}^{a} = -W_{\bm p' \bm p}^{a}$ is the asymmetric part of the elastic scattering probability. Here the brackets denote averaging over the directions of $\bm p$ at the Fermi circle, and the polar angles $\varphi_{\bm p}$ are reckoned from the $x$ axis.

\subsection{Photogalvanics in graphene at ESR}

The alignment of carrier' momenta also occurs in the ESR at direct optical transitions between spin-splitted energy levels. It is crucial that, as we show below, the momentum alignment in this case occurs in the opposite manner to the Drude-absorption case. Namely, a larger number of carriers is with momenta perpendicular to $\bm E$ (at energy $\varepsilon_p > \varepsilon_{\rm F}$), see magenta and blue solid arrows in the left side of Fig.~\ref{fig9}(b). Similarly to the mechanism described for Drude absorption above, these carriers also generate a photocurrent through skew scattering, but with a striking difference: the induced current reverses direction relative to the Drude-absorption contribution, see thick red arrow in the right side of Fig.~\ref{fig9}(b).

In order to get momentum alignment at ESR, an account of the classical magneto-dipole transitions is not enough. The alignment of carrier's momenta emerges if the absorption is due to an electric field of radiation enabled by the spin-orbit coupling -- electric dipole spin resonance (EDSR)~\cite{Rashba1960,Stier2023,Kumar2021,Denisov2024,Grigoryan2026}. EDSR in the heterostructures under study is caused by the Rashba spin-orbit interaction. At high energies exceeding by far the Zeeman and Rashba spin splittings, the Hamiltonian of the Rashba-coupled graphene in the presence of external magnetic field $\bm B$ for both valleys reads $\mathcal H = vp + \mathcal H_s$, where the spin part reads~\cite{Ilic2019,Golub2024}
\begin{equation}
\label{H_s}
 \mathcal H_s={1\over 2}g\mu_{\rm B} \bm B \cdot \bm s + \lambda_{\rm R} 
 %[\bm n_{\bm p}\times \bm s]_z 
 {[{\bm p}\times \bm s]_z \over p}. 
 %\equiv {\hbar \over 2}\bm s \cdot \bm \Omega.
\end{equation}
Here $g$ is the Land\'e factor, $\bm s$ is a vector of spin Pauli matrices, and $\lambda_{\rm R}$ is the Rashba spin-orbit constant. The perturbation $V = (ie/ \omega)\bm E \cdot \bm \nabla_{\bm p}\mathcal H_s$ resulting in the momentum alignment at EDSR due to the Rashba interaction. It has the following form 
\begin{equation}
\label{V_EDSR}
V = i{e \lambda_{\rm R}\over \omega p^3} (\bm s \cdot \bm p) [\bm E \times \bm p]_z.
\end{equation} 
We account for the Rashba splitting in the lowest order. It is included in the electron-photon interaction Eq.~\eqref{V_EDSR}, but ignored in the carrier energies. As a result, the absorption occurs at $\hbar\omega=g\mu_{\rm B}B$ for any direction of electron momentum $\bm p$.

First we consider magnetic field oriented perpendicular to the graphene plane ($\bm B \parallel z$). The matrix element~\eqref{V_EDSR}  is independent of the magnetic field strength and is given by
\begin{equation}
V_{+-}={e E\lambda_{\rm R}\over \omega p} [\bm E \times \bm p]_z.
\end{equation}

We calculate the photocurrent at ESR, $\bm j^{\rm ESR}$, introducing the distribution functions of photocarriers $f^\pm_{\bm p}$ in the $(\pm)$ spin subbands of the conduction band. They are given by
\begin{equation}
\label{align_ESR}
f^\pm_{\bm p} = \pm \tau_2^\pm G_{\bm p} \qty[f_0(\varepsilon_p-\hbar\omega/2)-f_0(\varepsilon_p+\hbar\omega/2)],
\end{equation}
where $\tau_2^\pm$ are the relaxation times $\tau_2$ for the subbands, and the EDSR transition rate per valley is given by the Fermi's golden rule
\begin{equation}
\label{G_Faraday}
	G_{\bm p} = {\pi \over \hbar} \qty({e\lambda_{\rm R}\abs{\bm E}\over \omega p})^2 [1-\cos{(2\varphi_{\bm p}-2\alpha)}] \delta_\gamma(g\mu_{\rm B}B_\perp-\hbar\omega).
\end{equation}
Here  $\delta_\gamma(x)=(\gamma/\pi)/(x^2+\gamma^2)$ with 
$\gamma= 1/\tau_s$, see Eq.~\eqref{tau_s}. 

Equation~\eqref{G_Faraday} describes momentum alignment in the upper conduction spin subband perpendicular to $\bm E$, $f^+_{\bm p} \propto -\cos{(2\varphi_{\bm p}-2\alpha)}$, Fig.~\ref{fig9}(b). The corrections $f^\pm_{\bm p}$ are even in momentum and do not contribute to the photocurrent. In order to get a nonzero current, we again take into account the skew scattering. The corresponding corrections to the distributions in the spin subbands, $\delta f^\pm_{\bm p}$, are found from the following kinetic equation
\begin{equation}
	{\delta f^\pm_{\bm p} \over \tau_1^\pm} + \sum_{\bm p'} W_{\bm p \bm p'}^{a}f^\pm_{\bm p'}=0.
\end{equation}
Here $\tau_1^\pm$ are the relaxation times of the first angular harmonics of the distribution function in the spin subbands. 

The photocurrent is given by
\begin{equation}
\label{Eq_j_ESR}
	\bm j^{\rm ESR} = 2e\sum_{\bm p, \pm}v{\bm p\over p}\delta f^\pm_{\bm p},
\end{equation}
where factor 2 accounts for the valley degeneracy. Calculation yields
\begin{equation}
	\label{j_ESR}
	j^{\rm ESR}_{x,y}(\bm B_\perp)= -ev \mathcal P_{\rm ESR}^\perp (\Xi_{c,s}P_{\rm L1}+\Lambda_{c,s}P_{\rm L2}) {(\tau_{\rm tr}\tau_{\rm al})'\over \tau_{\rm tr}},
\end{equation}
where $\Xi_{c,s}$ and $\Lambda_{c,s}$ are the skew scattering factors Eq.~\eqref{Xi_Lambda}, and the differentiation in Eq.~\eqref{j_ESR} stems from a difference in the products of the relaxation times in the spin subbands, which would otherwise cancel each other out in the photocurrent. The radiation power density absorbed at ESR
\begin{equation}
\mathcal P_{\rm ESR}=2\sum_{\bm p}\hbar\omega G_{\bm p}\qty[f_0(\varepsilon_p-\hbar\omega/2)-f_0(\varepsilon_p+\hbar\omega/2)]
\end{equation}
for $\hbar\omega, k_{\rm B}T \ll \varepsilon_{\rm F}$ and $\bm B_\perp$ directed out of plane is given by
\begin{equation}
\mathcal P_{\rm ESR}^\perp
= {2\pi\hbar n \over \varepsilon_{\rm F}} \qty({e E\lambda_{\rm R}\over  p_{\rm F}})^2  
\delta_\gamma(\hbar\omega-g\mu_{\rm B}B_\perp).
\end{equation}

Now we turn to the geometry $\bm B \parallel x$. In this configuration the square of the electro-dipole matrix element~\eqref{V_EDSR} is given by
\begin{equation}
\abs{V_{+-}}^2=\qty({e E\lambda_{\rm R}\over \omega p})^2 
\qty[\sin{(\varphi_{\bm p}-\alpha)} \sin{\varphi_{\bm p}}]^2.
\end{equation}
This results in a dichroism of absorption:
\begin{align}
&\mathcal P_{\rm ESR}^\parallel=\overline{\mathcal P}_{\rm ESR}\qty(1+{\cos{2\alpha} \over 2}),
\\
&\overline{\mathcal P}_{\rm ESR}={\pi\hbar n \over \varepsilon_{\rm F}}\qty({e E\lambda_{\rm R}\over  p_{\rm F}})^2  
\delta_\gamma(\hbar\omega-g\mu_{\rm B}B_\parallel).
\end{align}
We see that the absorbance for $\bm E \parallel \bm B_\parallel$ ($\alpha = 0$) is 3 times larger than for  $\bm E \perp \bm B_\parallel$ ($\alpha = 90^\circ$).

The LPGE current is calculated by Eq.~\eqref{Eq_j_ESR}.
Retaining the polarization-dependent part only and taking into account that $\expval{W^{a}_{\bm p' \bm p}}_{\varphi_{\bm p}}=0$, we obtain
\begin{equation}
	\label{j_ESR_B_par}
	j^{\rm ESR}_{x,y} = -ev \overline{\mathcal P}_{\rm ESR} \qty(\tilde{\Xi}_{c,s}P_{\rm L1}+\tilde{\Lambda}_{c,s}P_{\rm L2}) {(\tau_{\rm tr}\tau_{\rm al})'\over \tau_{\rm tr}}
\end{equation}
with the skew-scattering factors
\begin{align}
	\label{tilde_Xi_Lambda}
	&\tilde{\Xi}_{c,s} = \tau_\text{tr} \sum_{\bm p'}\left< \cos(\sin){\varphi_{\bm p}} W^{a}_{\bm p' \bm p} \qty(\cos{2\varphi_{\bm p'}}- {\cos{4\varphi_{\bm p'}}\over 2})\right>%_{\varphi_{\bm p}}
	,\nonumber
	\\
	&\tilde{\Lambda}_{c,s} = \tau_\text{tr} \sum_{\bm p'}\left< \cos(\sin){\varphi_{\bm p}} W^{a}_{\bm p' \bm p} \qty(\sin{2\varphi_{\bm p'}}- {\sin{4\varphi_{\bm p'}}\over 2})\right>%_{\varphi_{\bm p}}
	, 
\end{align}
where angular brackets mean averaging over $\varphi_{\bm p}$.
All four skew scattering factors $\tilde{\Xi}_{c,s}$, $\tilde{\Lambda}_{c,s}$ are nonzero and linearly independent due to the lowest possible ($C_1$) symmetry of the scattering potential.
They differ from the factors ${\Xi}_{c,s}$, ${\Lambda}_{c,s}$ for Drude absorption Eq.~\eqref{Xi_Lambda} by the last terms, $\propto \sin{4\varphi_{\bm p'}}$, $\cos{4\varphi_{\bm p'}}$. 
Although these terms enter with opposite signs, the common sign of  $\tilde{\Xi}_{c,s}$, $\tilde{\Lambda}_{c,s}$ most likely coincides with that of ${\Xi}_{c,s}$, ${\Lambda}_{c,s}$. 

For the geometries relevant to the experiment, where the photocurrent is measured either along or perpendicular to the electric field direction, we have $P_{\rm L2}=0$ and $P_{\rm L1}=1$ for $\bm E \parallel x$ or $P_{\rm L1}=-1$ for $\bm E \parallel y$. Then, from Eqs.~\eqref{j_Drude},~\eqref{j_ESR} and~\eqref{j_ESR_B_par}, we obtain for the total photocurrent $\bm j = \bm j^{\rm Drude} +\bm j^{\rm ESR}$
\begin{align}
	\label{j_xy}
	&j_{x,y}(\bm B_\perp) = ev P_{\rm L1} \qty(D_{x,y}\mathcal P_{\rm Drude} - R_{x,y}\mathcal P_{\rm ESR}^\perp), \\
	&j_{x,y}(\bm B_\parallel) = ev P_{\rm L1} \qty(D_{x,y}\mathcal P_{\rm Drude} - \tilde{R}_{x,y} \overline{\mathcal P}_{\rm ESR}). \nonumber
\end{align}
Here the second terms are only present in the vicinity of ESR, and
\begin{align}
	&D_{x,y} = {\left(\varepsilon_\text{F}^2 \tau_{\rm al} \Xi_{c,s}\right)'\over \varepsilon_\text{F}^2} 
	+
	(\Xi_{c,s}\mp\Lambda_{s,c}){\tau_{\rm al}\varepsilon_\text{F}\over \tau_{\rm tr}} \qty({\tau_{\rm tr} \over \varepsilon_\text{F}})',
	\nonumber
	\\
	&R_{x,y} = \Xi_{c,s} {(\tau_{\rm tr}\tau_{\rm al})'\over \tau_{\rm tr}}, \quad \tilde{R}_{x,y} = \tilde{\Xi}_{c,s} {(\tau_{\rm tr}\tau_{\rm al})'\over \tau_{\rm tr}},
\end{align}
where we took into account that $\omega\tau_{\rm tr,al} \ll 1$ for frequencies in the GHz range. These expressions show that the photocurrent direction is reversed in the ESR compared to the Drude background.

\subsection{ESR transition frequencies}
\label{transitionfrequencies}

Dirac electrons in our graphene samples near the $K_{\pm}$ valleys are described by the effective Hamiltonian
	\begin{multline}
		\label{H}
		\mathcal H^{K_\xi} = v (\xi \sigma_x p_x + \sigma_y p_y)
		+ \Delta_{\rm st}\,\sigma_z + 
		%\frac{\Delta}{2} s_z
		{1\over 2}g\mu_{\rm B} \bm B \cdot \bm s \\
		+\lambda_{\rm R} (\xi \sigma_x s_y - \sigma_y s_x) 
		+ \lambda_{\rm KM}\xi\sigma_z s_z + \lambda_{\rm VZ} \xi s_z,
	\end{multline}
	where $\xi=\pm1$ labels the two valleys, and $\boldsymbol{\sigma}$ and $\bm{s}$ are Pauli matrices acting in the sublattice and spin spaces, respectively. 	The second term describes the staggered sublattice potential, while the third term accounts for the Zeeman coupling to the external magnetic field $\bm{B}$, with the Land\'e factor $g\approx2$. 	The parameters $\lambda_{\rm R}$, $\lambda_{\rm KM}$, and $\lambda_{\rm VZ}$ denote the Rashba, Kane--Mele, and valley-Zeeman (Ising) spin-orbit coupling strengths, respectively.

	In pristine graphene, staggered potential, the Rashba and valley-Zeeman couplings vanish by symmetry. In graphene proximitized by, or encapsulated with, insulating layers such as hBN and WSe$_2$, however, these couplings can be induced by the proximity effect. The values range from 1-10 $\mu$eV to 1-10 meV~\cite{Zollner2021,Zollner2023}. Since the Fermi level in our samples is relatively large ($\varepsilon_{\rm F}>100$~meV), we neglect the staggered potential and the intrinsic Kane--Mele interaction, as their relative contributions to the spin texture decrease as $1/\varepsilon_{\rm F}$ with increasing Fermi energy. At the dopings relevant for our samples, the dominant spin-orbit interactions are therefore the Rashba and valley-Zeeman terms, see Eq.~\eqref{H_s}. 

Near the Fermi energy $\varepsilon_{\rm F} \gg g\mu_{\rm B}B_\perp,\lambda_{\rm R},\lambda_{\rm VZ}$, the energy spectrum is given by
\begin{equation}
	\label{band}
	\varepsilon_{c, \pm}^{K_\xi} \approx vp \pm \sqrt{(g\mu_{\rm B}B_\perp/2 + \xi \lambda_{\rm VZ})^2 +\lambda_{\rm R}^2}.
\end{equation}
Therefore, the spin splitting of the conduction subbands at large $p$ reads
\begin{equation}
	\label{Faraday}
	\Delta^{K_\pm}= \sqrt{(g\mu_{\rm B}B_\perp \pm 2\lambda_{\rm VZ})^2+ (2\lambda_{\rm R})^2}.
\end{equation}
However, the momentum scattering results in an effective averaging of the Rashba splitting over directions of $\bm p$. As a result, at $\lambda_{\rm R}\tau_{\rm tr}/\hbar \ll 1$ the Rashba splitting just slightly modify the resonance frequency. By contrast, the Zeeman and valley-Zeeman splittings are not affected by the momentum scattering and should exceed a much smaller spin relaxation rate to be seen in the ESR. Therefore at $g\mu_{\rm B}B_\perp, \lambda_{\rm VZ} \gg \hbar/\tau_s$, where $\tau_s$ is the spin relaxation time, the ESR in the Faraday (F) geometry takes place at different photon energies in the $K_{+}$ and $K_{-}$ valleys:
\begin{equation}	
\label{Faraday_no_Rashba}
	\hbar\omega_{\rm F}^{K_\pm}= \abs{g\mu_{\rm B}B_\perp \pm 2\lambda_{\rm VZ}} \qty[1+{2(\lambda_{\rm R}\tau_{\rm tr}/\hbar)^2 \over 1+(g\mu_{\rm B}B_\perp\tau_{\rm tr}/\hbar)^2}].
\end{equation}
The effect of Rashba splitting in this limit is restricted to a slight renormalization of the Larmor frequency similar to other 2D systems~\cite{Rakitskii2025}. By contrast, the valley-Zeeman splitting makes the resonance frequencies different in two valleys provided inter-valley scattering processes are ineffective. The difference between two ESR frequencies approximately equals to $4\abs{\lambda_{\rm VZ}}$.

The effect of the valley-Zeeman splitting is, however, smeared by the intervalley scattering, in a manner analogous to the magnetotransport situation~\cite{Golub2026}. The ESR frequencies are given by
\begin{equation}
\label{Faraday_iv_scatt}
\hbar\omega_{\rm F}^{K_\pm} = g\mu_{\rm B}B_\perp \pm {\rm Re}\sqrt{(2\lambda_{\rm VZ})^2-(\hbar/2\tau_{iv})^2},
\end{equation}
where $\tau_{iv}$ is the intervalley scattering time. This expression shows that the ESR frequency has a single value $\omega_{\rm F} = g\mu_{\rm B}B_\perp/\hbar$ when $4\lambda_{\rm VZ}\tau_{iv}/\hbar <1$.

In the Voigt (V) geometry, where magnetic field $\bm B_\parallel$ is parallel to the graphene plane $(xy)$, diagonalization of the Hamiltonian~\eqref{H} with the last term $\propto \bm B_\parallel \cdot \bm s_\parallel$ gives the following energies for the conduction spin subbands:
\begin{align}
	\label{band_V}
&	\varepsilon_{c, \pm}^{K_\xi} \approx vp 
	\\ 
& \pm \sqrt{(g\mu_{\rm B}B_\parallel/2)^2 + \lambda_{\rm VZ}^2+ \lambda_{\rm R}^2 - \lambda_{\rm R}g\mu_{\rm B} {[\bm B_\parallel \times \bm p]_z\over p} }. \nonumber
\end{align}
Therefore, the ESR for electrons with the momentum $\bm p$ takes place at the same frequency for both valleys.

If the Rashba spin splitting is larger than the broadening due to momentum scattering, $\abs{\lambda_{\rm R}} \gg \hbar/\tau_{\rm tr}$, it affects the resonance frequency. The ESR frequencies lie in the band between the $\omega^{\rm max}_{\rm V}$ and $\omega^{\rm min}_{\rm V}$ which are given by
\begin{equation}
\label{resVoigt1}
\hbar\omega^{\rm max/min}_{\rm V}=\sqrt{\qty(\abs{g\mu_{\rm B}B_\parallel} \pm 2\abs{\lambda_{\rm R}})^2 + (2\lambda_{\rm VZ})^2 }.
\end{equation}
The density of states diverges at the borders of this band~\cite{Glenn2012}, therefore, the ESR has van Hove singularities at $\omega=\omega^{\rm min}_{\rm V}$ and $\omega=\omega^{\rm max}_{\rm V}$.

However in the opposite limit $\abs{\lambda_{\rm R}}\tau_{\rm tr}  /\hbar \ll 1$ relevant to the experiment, analysis of the spin dynamics in the presence of frequent momentum scattering shows that the Rashba splitting gives a small correction to the spin precession frequency~\cite{Larionov2008,Yagodkin2025}. This happens due to motional narrowing similar to the Faraday configuration. As a result, we obtain for the resonance frequency
%However in the opposite limit $\abs{\lambda_{\rm R}}\tau_{\rm tr}  /\hbar \ll 1$ relevant to the experiment, due to the reasons addressed above discussing the ESR in the Faraday configuration, we obtain a single resonance frequency
\begin{equation}
\label{resVoigt2}
\hbar\omega_{\rm V}=\sqrt{(g\mu_{\rm B}B_\parallel )^2 + (2\lambda_{\rm VZ})^2 }.
\end{equation}
Furthermore, similar to the Faraday geometry, intensive inter-valley scattering suppresses effect of the valley-Zeeman splitting on the ESR frequency. If $\lambda_{\rm VZ}\tau_{iv}/\hbar \ll 1$, there are just small corrections quadratic in $\lambda_{\rm VZ}$.

To summarize this part, in the limit of intensive intra- and intervalley scattering, the resonance frequency reads 
\begin{equation}
\hbar\omega=g\mu_{\rm B}B
\end{equation}
for both valleys and for any magnetic field orientation.

\section{Discussion}
\label{Discussion}

\subsection{LPGE in graphene}

All of the experimental results discussed above have several features in common: (i) the ESR-driven and non-resonant Drude absorption-driven photocurrents have consistently opposite signs for all structures, radiation frequencies, and carrier types/densities (Figs.~\ref{fig2}, \ref{fig3}, \ref{fig4}, \ref{fig6}, \ref{fig7}, and~\ref{fig8}); (ii) the resonant and non-resonant photocurrents are even in a magnetic field $\bm B$ (Figs.~\ref{fig3},  and~\ref{fig8}), (iii) reverse their sign in the vicinity of the CNP (Figs.~\ref{fig4}, and \ref{fig7}); and (iv) the resonances are well-fitted by a Lorentzian function, and their positions in the magnetic field depend almost linearly on the radiation frequency with different intercepts at $B=0$ (Figs.~\ref{fig2}, \ref{fig3}, \ref{fig4}, \ref{fig6}, and \ref{fig7}). The fact that the even in magnetic field contribution of both the non-resonance and resonance responses dominates, demonstrates that the magneto-photogalvanic effect~\cite{Belkov2008,Drexler2013}, which is odd in magnetic field, is negligible. Furthermore, as noticed above we can exclude the edge photogalvanic currents, because in the discussed experiments we used radiation electric field oriented along or perpendicular to the sample edges, for which the edge LPGE vanishes~\cite{Candussio2021,Candussio2021a}.

As discussed above, in general, three microscopic mechanisms can contribute to the nonresonant LPGE current, namely skew scattering, side jump and Berry curvature for $\bm j \perp \bm E$~\cite{Moldavskaya2026}. In the geometry used in our experiments, $\bm j \parallel \bm E$, the latter one is absent. The photocurrent caused by the ESR may have in addition to  the skew scattering, two other contributions arising at the moment of absorption, the ballistic and shift mechanisms of the LPGE.
%~\cite{Golub2020,Moench2025}. 
The latter one is caused by the shift of the wave package at the direct optical transition, and, generally speaking is an analog to the side jump at indirect intraband optical transitions. We believe that observed opposite signs for Drude and ESR photocurrents reveal that the skew scattering is a dominant mechanism for the current formation in both cases. As discussed above, the opposite signs are caused by the 90 degree difference in the alignment for direct and indirect transitions resulting in systematically opposite signs of the related photocurrents, see Figs.~\ref{fig9}(a) and (b). The skew scattering mechanism described in the previous section, results in the $dc$-current whose direction can only be caused by either variation of Drude absorption and transport parameters, like mobility, or formation of new magnetic-field induced absorption channels, such as ESR. Note that for the Drude absorption the photocurrent is also generated at zero magnetic field, see Figs.~\ref{fig3},  and~\ref{fig8}.
 
The observed reversal of the photocurrent sign close to the CNP is also in agreement with the theory of the LPGE: Eqs.~\eqref{j_xy} shows that the current $\bm j$ is proportional to the odd power of the carrier charge, thus changing from positively charged holes to negatively charged electrons results in the inversion of the current direction, see Figs.~\ref{fig4} and \ref{fig7}. Note that in our experiments we detected the photovoltage $U$, which is coupled to the photocurrent $J \propto j$ as $J = U / R$, where $R$ is the sample resistance in the direction in which the photoresponse is picked-up. Figs.~\ref{fig4}(d) and \ref{fig7}(d) present the data obtained in a wide range of the gate voltages, i.e., in the range in which the resistance strongly varies (see gray line in the figures). In the vicinity of the CNP the signal depends almost linear on the gate voltage, i.e., carrier density. At higher negative/positive gate voltages, however the photovoltage  $U$ tends to saturate or even decreases. This is mostly due to the drastic decrease of the sample resistance, which compensates the growth of the current magnitude caused by the increase of the corresponding absorption coefficients.

\subsection{ESR width and multiple resonances}

Finally, we discuss the shapes and positions of the ESR resonances. Our experiments, which were performed at high Fermi energies (up to 60~meV in MLG and 200~meV in WSe$_2$/MLG), demonstrate that the ESR-driven photogalvanic current exhibits two resonances. Furthermore, in a number of our experiments a third weak resonance signal has also been detected.  These resonances scale linearly with the radiation frequency and have full widths at half maximum of about $0.05–0.6$~T, corresponding to spin relaxation times of approximately $18–130$~ps. As mentioned previously, the detected resonances are excited in the graphene bulk because the edge LPGE vanishes when the electric field is oriented along or perpendicular to the edges, which is the geometry used in our experiments. For both geometries, the resonance frequencies can be well fitted by the empirical relation
\begin{equation}
	\label{fit}
	\hbar\omega = g\mu_{\rm B}B + \Delta.
\end{equation}
For both the Faraday and Voigt geometries we obtained the Land\'e factor for both kinds of structures  $g\approx 1.99$. The intercepts at $B=0$ of the first, second and third resonances are also very close for both structures and geometries. They are $\Delta \approx -5\div -7$~GHz (21-29~$\mu$eV), zero and $10.2$~GHz ($42.2~\mu$eV), respectively. Comparison of our results for the out- and in-plane magnetic fields show that the resonances and their characteristics are nearly isotropic, see Figs.~\ref{fig2}(c), \ref{fig3}(c), \ref{fig6}(c) and~\ref{fig8}(c). Note that a fourth resonance, $R_4^{\rm V},$ was clearly detected in the MLG structures subjected to an in-plane magnetic field, as shown in  Fig.~\ref{figA3} in  Appendix~\ref{B}.

As our analysis performed in Sec.~\ref{transitionfrequencies} shows, that two resonances ($R_1$ and $R_3$) with non-zero intercepts at $B=0$  are indeed possible even for high Fermi energies. In the Faraday geometry $R_1^{\rm F}$ and $R_3^{\rm F}$ can be  caused by the valley-Zeeman splitting and are given by Eq.~\eqref{Faraday} for $\hbar\omega = \Delta^{K_\pm}$. In the Voigt geometry  resonances $R_1^{\rm V}$ and $R_3^{\rm V}$ may be due to Rashba spin-orbit splitting. In this case, considering the van Hove singularities, their positions are given by $\omega^{\rm min}_{\rm V}$ and $\omega^{\rm max}_{\rm V}$,  see Eq.~\eqref{resVoigt1}. However, as we show below, both   Eqs.~\eqref{Faraday} and~\eqref{resVoigt1} are not applicable for the studied samples.

The effect of the Rashba splitting, needed for existence of two resonances $R_1^{\rm V}$ and $R_3^{\rm V}$ in the Voigt geometry, is suppressed due to not sufficiently high mobility of our structures. Indeed, in the systems with the Rashba spin splitting, the spin relaxation times are defined by the Dyakonov-Perel mechanism, which yields $\tau_s^{-1} \sim (\lambda_{\rm R}  /\hbar)^2\tau_{\rm tr}$. The transport relaxation times are determined from the mobility measurements. Figures~\ref{fig1_Gr_setup}(d) and~\ref{fig5}(d) show that mobility in our MLG  is of the order of $0.5\times10^6$ and in WSe$_2$/MLG of $10^4$~cm$^2$/Vs. Consequently for the carrier densities of the order of 10$^{11}$~cm$^{-2}$ we obtain transport relaxation times 1.8~ps and~0.037~ps for MLG and WSe$_2$/MLG, respectively. Taking the experimental values $\tau_s$ in the the range from 18 to 130~ps for both structures, we obtain the upper bound from $\lambda_{\rm R}<0.04$~meV to $\lambda_{\rm R}<0.1$~meV for MLG and $\lambda_{\rm R}<0.3$~meV to $\lambda_{\rm R}<0.8$~meV for WSe$_2$/MLG. All these values are much smaller than $\hbar/\tau_{\rm tr}$, and we conclude that the Rashba splitting in this case is smeared by disorder.  The upper bound for the Rashba spin splitting $2\lambda_{\rm R}$ is in agreement with studies of the Shubnikov-de-Haas oscillations and  weak antilocalization in different TMDC/MLG structures. The  enhancement of spin-orbit interaction (SOI) in comparison with MLG has been clearly confirmed by several groups, which have shown that in WSe$_2$/MLG and WS$_2$/MLG  structures the strength of  SOI in graphene   is  strongly enhanced up to several meV~\cite{Wang2015,Yang2016,Voelkl2017,Wakamura2018,Zihlmann2018,Amann2022,Rockinger2026}. While our data can not confirm such an enhancement, the estimated upper limit of the Rashba splitting has the same order of magnitude.  Note that Figs.~\ref{fig4}, \ref{fig6} and~\ref{fig8}  demonstrate that  the resonance positions are almost independent of the applied gate voltage. Summarizing this paragraph, for  $\abs{\lambda_{\rm R}}\tau_{\rm tr}  /\hbar \ll 1$ relevant to the experiment we obtain a single resonance frequency given by Eq.~\eqref{resVoigt2}, where the Rashba splitting does not enter. 
 
 Using $\hbar\omega_{\rm F}^{K_\pm}= \abs{g\mu_{\rm B}B_\perp \pm 2\lambda_{\rm VZ}}$ followed from Eq.~\eqref{Faraday_no_Rashba} for description of our results, we may attribute the splitting $\Delta$ detected for the resonances $R_1^{\rm F}$ and $R_3^{\rm F}$ with the valley-Zeeman splitting  and conclude that $\lambda_{\rm VZ} = 5$~$\mu$eV. However, this analysis fails for the Voigt geometry, where we have only one resonance frequency, Eq.~\eqref{resVoigt2}, even in the presence of the valley-Zeeman splitting. Experimental results demonstrating almost identical responses for Faraday and Voigt geometries exclude the valley-Zeeman splitting as a source for the ESR peak's splitting. An absence of a considerable valley-Zeeman interaction induced ESR peak splitting in the experiments can be explained by effective inter-valley scattering which smears the effect of the valley-Zeeman splitting, see Eq.~\eqref{Faraday_iv_scatt}.
 
 The valley-Zeeman and the Rashba splitting as a cause of the two resonances can also be excluded considering the  resonance widths.  The extracted spin relaxation times $\tau_s$ ranging from  18 to 130~ps are comparable to those reported in the first generation of graphene spin-transport devices~\cite{Tombros2007}. In contrast, state-of-the-art spin transport measurements typically yield spin relaxation times, even at room temperature, in the range of $1$--$10$~ns~\cite{Singh2016,Droegeler2016,Guimaraes2014,Yang2011a}. The origin of the relatively broad ESR resonances remains unclear, particularly since they are detected at low temperatures and their widths show no pronounced dependence on the sample mobility. The longer spin lifetimes reported in modern transport experiments are generally attributed to improved material quality, including hBN encapsulation, reduced disorder, and minimized contact-induced spin relaxation. The shorter ESR-derived lifetimes may result from additional dephasing mechanisms inherent to resonance measurements, such as inhomogeneous broadening. The absence of systematic trends in the observed linewidths precludes drawing definitive conclusions about the microscopic origin of the ESR broadening.

\begin{figure}
	\centering
	\includegraphics[width=\linewidth]{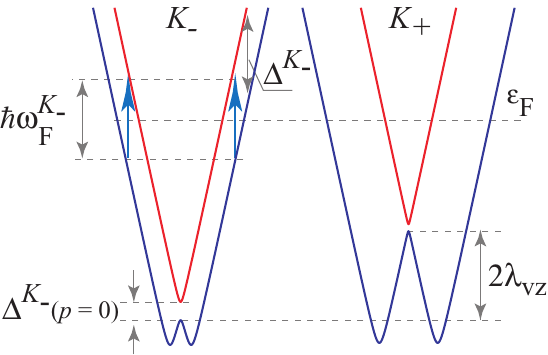}
	\caption{ Illustration of the graphene band structure obtained from  the Hamiltonian~\eqref{H} for $\bm B_\perp$ perpendicular to the graphene plane and  non-zero $\lambda_{\rm VZ}$ and $\lambda_{\rm R}$. The band splittings at zero and large momenta, $\Delta^{K_-}(p=0)$ and $\Delta^{K_-}$, are given by Eqs.~\eqref{zeroDelta} and  \eqref{Faraday}, respectively. The blue upward arrows illustrate possible direct ESR transitions in the Faraday geometry that occur at a high Fermi energy position, $\varepsilon_{\rm F}$. According to the conservation laws, these transitions are allowed for the $K_{-}$ valley but not the $K_{+}$ valley in a constant magnetic field. We assume that both $B_\perp$ and $\lambda_{\rm VZ}$ are positive, and the intervalley scattering is ineffective.
	}
	\label{fig10}
\end{figure} 

Analyzing the literature on the ESR in graphene structures in magnetic fields perpendicular to the structure plane~\cite{Mani2012,Lyon2017,Sichau2019,Singh2020,Anlauf2021,Morissette2023, Sharma2022,Dinar2025}, we found that the resonances with the same positions (up to renormalization of frequencies and magnetic fields) and widths as our first, second and third resonances, and even the fourth one, were frequently reported in the Faraday geometry, see gray areas in Figs.~\ref{fig2}(c), and \ref{fig6}(c). The multiple (up to four) resonances have been attributed to the interband transitions accounting for the zero field Kane-Mele  and sublattice splittings~\cite{Sichau2019,Singh2020,Anlauf2021, Sharma2022,Dinar2025}. At high Fermi energies and low temperatures, however, these arguments can not be applied for our work. Indeed, in the analyzed experiments applying radiation with $\hbar \omega \ll \varepsilon_{\rm F}$ (photon energies in the range from 0.19 to 0.3~meV and the Fermi energy up to 60~meV in MLG and 200~meV in WSe$_2$/MLG) the initial and final states of the  optical transitions are far from the Dirac points in both valleys, and interband transitions are forbidden by the Pauli principle, see Fig.~\ref{fig10}. Therefore, at high Fermi energies as in our samples, the role of the Kane-Mele splitting as well as of the staggered potential, is negligible, as it is pointed out in Sec.~\ref{transitionfrequencies}. For the Voigt geometry, we found one work, Ref.~\cite{Morissette2023}, in which twisted graphene has been studied. The position of all four resonances presented in this work are the same as in our study. The resonances observed in Ref.~\cite{Morissette2023}, as in our work, are fully isotropic in respect to the magnetic field orientation, and were attributed to the formation of the Moir\'e superlattice.  Furthermore, the resonance widths are similar in all  studies performed by different groups on a very different graphene structures including MLG, bilayer graphene, trilayer graphene, ratchet structures on MLG and bilayer graphene, MLG with periodic antidots, TMDC/MLG with mobilities ranging from 500 to  $0.5\times10^6$~cm$^2$/Vs (used in the present work)~\cite{Mani2012,Lyon2017,Sichau2019,Singh2020,Anlauf2021,Bray2022,Sharma2022,Hild2024a,Morissette2023,Dinar2025}. The corresponding spin relaxation times in these studies are also rather short scaling from   12 to 110~ps~\cite{Lyon2017,Singh2020,Anlauf2021,Sharma2022,Bray2022,Dinar2025}, which, as addressed in previous paragraph, is surprising.

Summarizing this part, we emphasize that the origin of the splitting and widths of ESR resonances at high Fermi energies relevant to our work remain puzzling. Its understanding requires further study, which is out of scope of our paper aimed at the ESR driven photogalvanic effect.

\section{Summary and Outlook}  

In summary, we have observed an electron spin resonance driven linear photogalvanic response in unbiased monolayer graphene and WSe$_2$/graphene heterostructures. Resonant photovoltage features are observed in both Faraday and Voigt geometries over a broad range of radiation frequencies. The resonance fields vary linearly with frequency, while the resonant amplitudes reverse sign across the charge-neutrality point and have the opposite sign to the nonresonant Drude background. These common features are described by a microscopic theory based on radiation-induced momentum alignment and skew scattering. More generally, our results establish resonant photogalvanic detection as an unbiased probe of spin resonance in micron-scale two-dimensional systems. Further measurements over a broader range of experimental conditions, together with additional theoretical analysis, will be needed to clarify the microscopic origin of the observed multiple ESR resonances.

\section{Acknowledgments}
The authors thank M.M. Glazov and J. Wunderlich for fruitful discussions. The support of the Deutsche Forschungsgemeinschaft  (DFG, German Research Foundation) via project 572024380 (WU 883/4-1) and the European Union through the ERC-ADVANCED grant TERAPLASM No. 101053716 is gratefully acknowledged.  Views and opinions expressed are, however, those of the author(s) only and do not necessarily reflect those of the European Union or the European Research Council Executive Agency. Neither the European Union nor the granting authority can be held responsible for them. Work of L.E.G., J.F. and A.T. was funded by the German Research Foundation (DFG) as part of the German Excellence Strategy -- EXC3112/1 -- 533767171 (Center for Chiral Electronics).  This work was also supported by the French Agence National pour la Recherche and by the France 2030 program through Equipex+ HYBAT project (ANR-21-ESRE-0026), and PEPR Electronique, COMPTERA project (ANR-22-PEEL-0003). A.T. acknowledges funding by the Deutsche Forschungsgemeinschaft (DFG, German Research Foundation) – Project-ID 535253440 – SPP 2244 2DMP (TU 149/21-1), and  funding from the European Union Graphene Flagship project 2D Materials of Future 2DSPIN-TECH (No. 101135853). M.M. and J.E. acknowledge funding by the Deutsche Forschungsgemeinschaft (DFG, German Research Foundation) – Project-ID 314695032 – SFB 1277 (subproject A09). K.W. and T.T. acknowledge support from the JSPS KAKENHI (Grant Numbers 21H05233 and 23H02052) , the CREST (JPMJCR24A5), JST and World Premier International Research Center Initiative (WPI), MEXT, Japan.

\appendix

\section{Spin splittings at zero momentum}

In the magnetic field $\bm B_\perp$ perpendicular to the graphene plane, the energies of electrons in the two spin subbands of the  conduction band ($c,\pm$) of the valley $K_\xi$ at $p=0$ read 
\begin{align}
	\label{Zero_p}
&	\varepsilon_{c, +}^{K_\xi} (p=0) = \sqrt{(g\mu_{\rm B}B_\perp/2 + \xi \lambda_{\rm VZ})^2 +(2\lambda_{\rm R})^2},
\\
&	\varepsilon_{c, -}^{K_\xi} (p=0) = \abs{g\mu_{\rm B}B_\perp/2 + \xi \lambda_{\rm VZ}}.
\end{align}
One can see that the energies of the lower spin subbands in two valleys, $\varepsilon_{c, -}^{K_+}$ and $\varepsilon_{c, -}^{K_-}$ differ at $p=0$ by $2\lambda_{\rm VZ}$, see Fig.~\ref{fig10}. The spin splitting at $p=0$ is given by
\begin{equation}
	\label{zeroDelta}
\Delta^{K_\pm}(p=0)= \varepsilon_{c, +}^{K_\xi} (p=0)-\varepsilon_{c, -}^{K_\xi} (p=0).
\end{equation}

For the magnetic field $\bm B_\parallel$ parallel to the graphene plane $(xy)$, diagonalization of the Hamiltonian~\eqref{H} with the last term $\propto \bm B_\parallel \cdot \bm s_\parallel$ gives equal energies for both valleys. For the conduction spin subbands at $p=0$ we have
\begin{equation}
\varepsilon_{c, \pm}(p=0)= \sqrt{(g\mu_{\rm B}B_\parallel/2)^2 +2\lambda_{\rm R}^2 + \lambda_{\rm VZ}^2 \pm U},
\end{equation}
where 
\begin{equation}
U=\lambda_{\rm R}\sqrt{(g\mu_{\rm B}B_\parallel)^2 +4\lambda_{\rm R}^2}.
\end{equation}

\section{Additional data with focus on the third resonance}
\label{B}

\begin{figure}
	\centering
	\includegraphics[width=\linewidth]{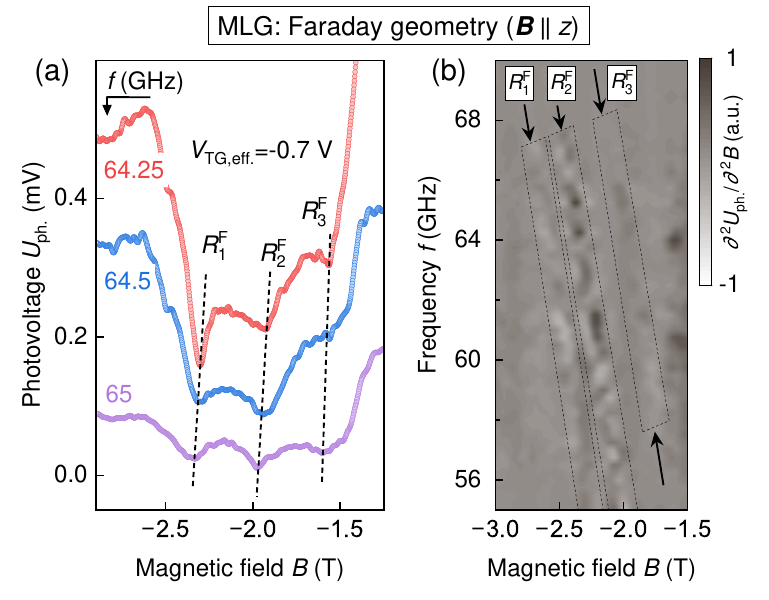}
	\caption{(a):~Photovoltage as a function of the magnetic field  measured in the Faraday geometry $\bm B\parallel z$ in the MLG device. The data were obtained for the radiation with a frequency of 64.25~GHz~(red curve), 64.5~GHz~(blue curve) and  65~GHz~(violet curve).  The radiation electric field vector is  parallel to the $x$-axis,  $V_{\rm BG,eff}$ = -0.7~V and a temperature of $T=1.9$~K. Black dashed lines represent the resonance positions, labeled as $R_1^{\rm F}$, $R_2^{\rm F}$ and $R_3^{\rm F}$. (b):~Second derivative of the photovoltage,  $\partial^2 U_{\rm ph} /\partial^2 B$ as a function of the in-plain  magnetic field and the radiation frequency. The data are presented for $V_{\rm BG,eff}$=-0.7~V and $T=1.9$~K. The black dashed regions and arrows in the middle of the map are used to highlight the ESR positions.}
	\label{figA1}
\end{figure}

\begin{figure}
	\centering
	\includegraphics[width=\linewidth]{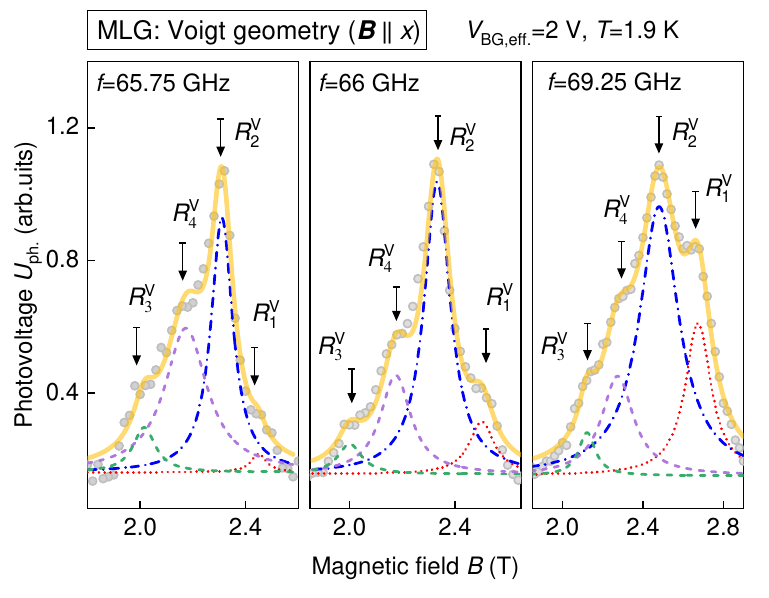}
	\caption{Photovoltage as a function of the magnetic field  measured in the MLG device in the Voigt configuration with the in-plane magnetic field $\bm B \parallel x$. The data were obtained for  V$_{\rm BG,eff}=2$~V, temperature of $T=1.7$~K  and the radiation with different frequencies given in the left top corner of the panels. The experimental data (gray  circles) are fitted by a sum of four Lorentzian functions (yellow solid curve), with the resonance positions indicated by downward arrows labeled $R_1^{\rm V}$, $R_2^{\rm V}$ and $R_3^{\rm V}$ and $R_4^{\rm V}$. The corresponding individual Lorentzian are shown by the magenta dotted, blue dashed-dotted, green dashed  and violet dashed  curves, respectively. 
	%The times used for the fits are calculated using Eq.~\eqref{tau_s} and are 33, 84, and 120.5~ps for the resonances $R_1^{\rm F}$, $R_2^{\rm F}$, and $R_3^{\rm F}$, respectively. 
	 Vertical arrows indicate the positions of the resonances $R_1^{\rm V}$, $R_2^{\rm V}$, $R_3^{\rm V}$ and  $R_4^{\rm V}$.  
	}
	\label{figA3}
\end{figure}

\begin{figure*}
	\centering
	\includegraphics[width=\linewidth]{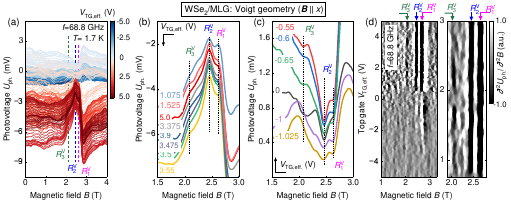}
	\caption{(a):~Photovoltage as a function of the magnetic field  measured in the WSe$_2$/MLG device in the Voigt configuration with the in-plane magnetic field $\bm B \parallel x$. The data were obtained for the radiation with a frequency of 68.8~GHz,  an electric field vector parallel to the $y$-axis,  a temperature of $T=1.7$~K and back gate voltage 10~V.  The  effective gate voltages was varied  from V$_{\rm BG,eff}=-5$ to 5~V, see the color bar on the right.  Vertical dashed lines indicate the positions of the resonances $R_1^{\rm V}$, $R_2^{\rm V}$ and $R_3^{\rm V}$.  Panels (b) and (c) show  a few  curves for positive V$_{\rm TG,eff}$, ranging from 1.05 to 5.0~V and for negative $V_{\rm TG,eff}$, ranging from 0 to -1.025~V, respectively. The color map shows the second derivative of the photovoltage,   $\partial^2 U_{\rm ph} /\partial^2 B$, with respect to the magnetic field and the effective top gate voltage. The data are presented for the radiation frequency $f=68.8$~GHz and $T=1.7$~K. The ESR positions are highlighted by arrows and labeled as $R_1^{\rm V}$,   $R_2^{\rm V}$   and $R_3^{\rm V}$.
	}
	\label{figA2}
\end{figure*}

The third resonances ($R_3^{\rm F}$ and $R_3^{\rm V}$) detected in the MLG and WSe$_2$/MLG structures, which are discussed in the main text, were weaker than the first and second resonances. However, they are still clearly detected in several of our experiments on both types of graphene devices. Figures~\ref{figA1}, \ref{figA3} and \ref{figA2} present the results of the experiments in which the third resonance was most clearly detected. Figure~\ref{figA3} also presents the data on MLG devices in which the fourth resonance $R_4^{\rm V}$ was clearly detected.

%\begin{figure*}
%	\centering
%	\includegraphics[width=\linewidth]{Fig_SM_TMDs.png}
%	\caption{TMDC/Graphene}
%	\label{figA3}
%\end{figure*} 

%\begin{figure*}
%	\centering
%	\includegraphics[width=\linewidth]{Fig_SM_TMDs.pdf}
%	\caption{Graphene}
%	\label{figA4}
%\end{figure*} 

\bibliography{all_lib_ESR.bib}

\end{document}